\documentclass[sigconf]{acmart}

\copyrightyear{2026}
\acmYear{2026}
\setcopyright{cc}
\setcctype{by}
\acmConference[SIGCSE TS 2026] {Proceedings of the 57th ACM Technical Symposium on Computer Science Education V.1}{February 18--21, 2026}{St. Louis, MO, USA.}
\acmBooktitle{Proceedings of the 57th ACM Technical Symposium on Computer Science Education V.1 (SIGCSE TS 2026), February 18--21, 2026, St. Louis, MO, USA}
\acmISBN{979-8-4007-2256-1/2026/02}
\acmDOI{10.1145/3770762.3772648}

\usepackage{tabularx}
\begin{document}

\title[Systems for Scaling Accessibility Efforts in Large Computing Courses]{Systems for Scaling Accessibility Efforts\\in Large Computing Courses}

\author{Ritesh Kanchi}
\email{rkanchi@uw.edu}
\orcid{0009-0006-7978-0821}
\affiliation{%
  \institution{University of Washington}
  \city{Seattle}
  \state{Washington}
  \country{USA}
}

\authornote{The author was affiliated with the University of Washington when this work was conducted and is currently affiliated with Harvard University.}

\author{Miya Natsuhara}
\email{mnats@cs.washington.edu}
\orcid{0009-0008-7464-3670}
\affiliation{%
  \institution{University of Washington}
  \city{Seattle}
  \state{Washington}
  \country{USA}
}

\author{Matt X. Wang}
\email{mxw@cs.washington.edu}
\orcid{0009-0003-0708-9378}
\affiliation{%
  \institution{University of Washington}
  \city{Seattle}
  \state{Washington}
  \country{USA}
}

\renewcommand{\shortauthors}{Ritesh Kanchi, Miya Natsuhara, \& Matt X. Wang}


\begin{abstract}
It is critically important to make computing courses accessible for disabled students. This is particularly challenging in large computing courses, which face unique challenges due to the sheer scale of course content and staff. In this experience report, we share our attempts to scale accessibility efforts for a large university-level introductory programming course sequence, with over 3500 enrolled students and 100 teaching assistants (TAs) per year. First, we introduce our approach to auditing and remediating course materials by systematically identifying and resolving accessibility issues. However, remediating content post-hoc is purely reactive and scales poorly. We then discuss two approaches to systems that enable proactive accessibility work. We developed technical systems to manage remediation complexity at scale: redesigning other course content to be web-first and accessible by default, providing alternate accessible views for existing course content, and writing automated tests to receive instant feedback on a subset of accessibility issues. Separately, we established human systems to empower both course staff and students in accessibility best practices: developing and running various TA-targeted accessibility trainings, establishing course-wide accessibility norms, and integrating accessibility topics into core course curriculum. Preliminary qualitative feedback from both staff and students shows increased engagement in accessibility work and accessible technologies. We close by discussing limitations and lessons learned from our work, with advice for others developing similar auditing, remediation, technical, or human systems.
\end{abstract}



\received{3 July 2025}
\received[accepted]{15 September 2025}

\maketitle

\section{Introduction}

Many conventions used to teach computer science (CS) exclude disabled students, particularly when visual metaphors, complex diagrams, or interactive interfaces lack accessible alternatives \cite{burden-of-survival, including-blind-people}. Further barriers arise from how course materials are distributed, how content is presented, and how students interact with the course itself \cite{access-differential}. Computing educators have both the opportunity and the responsibility to address these accessibility challenges \cite{accesscsforall}.

These efforts are critical for supporting disabled students. In 2020 NPSAS data, 21\% of U.S. undergraduates reported having a disability \cite{GAOHigherEd}. As of 2024, only 4.6\% of North American CS undergraduates received disability accommodations \cite{taulbee}, while 27.8\% of CS departments report offering no accommodations at all \cite{taulbee}.

In this experience report, we share our attempts to scale accessibility efforts in a large, university-level introductory programming course sequence. Each year, the sequence serves over 3500 enrolled students and employs more than 100 teaching assistants (TAs) across three courses.

We begin by reviewing existing policies and support for disabled students, accessibility education in computing curricula, and efforts to train TAs in teaching accessibly. Next, we describe our own interventions, organized across four interrelated systems:

\begin{enumerate}
    \item \textbf{Auditing} and \textbf{Remediation systems} are tightly coupled: audits systematically identify where accessibility barriers exist in course materials, while remediation describes structured methods and tools to resolve these issues.
    \item \textbf{Technical systems} manage the complexity of remediation at scale through redesigned course materials, alternative views, and frequent automated accessibility tests.
    \item \textbf{Human systems} empower TAs and students to adopt accessibility best practices through hands-on trainings, course-wide accessibility norms, and direct curricular integration.
\end{enumerate}

Through these systems, we share lessons learned from implementing and scaling accessibility efforts in large computing courses. By combining technical infrastructure with human-centered strategies, we highlight practical approaches that others can adapt or build upon to improve accessibility in computing education at scale. To support computing educators in similar initiatives, we have made supporting materials available online.\footnote{\url{https://g.riteshkanchi.com/sigcse26-systems-for-scaling-accessibility/}}

\section{Background}
Existing efforts to support accessibility are longstanding, spanning both instructional practices and institutional policies. We begin by reviewing the laws and institutional support that shape accessibility accommodations. We then explore how accessibility is taught in computing curricula and conclude by examining how TAs are---or are not---trained to teach accessibly in their courses.

\subsection{Legal Policies and Institutional Support}
Title II of the Americans with Disabilities Act (ADA) requires states and local governments in the U.S.---including public universities and community colleges---to ensure that services, programs, and activities are accessible to disabled individuals \cite{ada-fact-sheet}. Other legislation, such as the Accessibility for Ontarians with Disabilities Act \cite{aoda} and the European Accessibility Act \cite{eaa} also have similar compliance requirements. However, applying these standards at scale is challenging in higher education, due to the diversity of curricula and varied levels of accessibility expertise \cite{tac-disability-accessibility}.

Institutional accessibility support is distributed across services such as access technology centers---where students explore and engage with assistive technologies (ATs)---and disability services offices, which manage accommodations and communicate student needs. These services are critical in supporting disabled students, but are frequently under-resourced and complex to navigate \cite{disability-services-grad-support, navigating-complexity}. Institutional support focuses on individual needs rather than proactive educational design, limiting its scale and effectiveness.

\subsection{Teaching Accessibility in Computing}
Accessibility is often underemphasized in computing education. When taught, it often appears in Human-Computer Interaction or design courses \cite{a-systematic-analysis-of-accessibility}; outside of these courses, accessibility is rarely treated as a core topic and few resources exist \cite{who-teaches-accessibility,tac-introduction}. Prior research explores how programming assignments in introductory programming courses can highlight the technical aspects of accessibility \cite{infusing-accessibility-into-programming}. Beyond these introductory programming courses, accessibility has also been integrated into subfields ranging from accessible visualization design for blind and low vision (BLV) individuals to identifying and addressing dataset bias impacting disabled communities \cite{beyond-hci, mapping-accessibility, teaching-about-accessibility}. This broad applicability underscores the need to integrate accessibility not only earlier in students' education, but more consistently across computing subfields.

\subsection{Training Teaching Assistants to Teach Accessibly}
There is a distinction between \textit{teaching accessibility} (to explicitly teach accessibility-related concepts) and \textit{teaching accessibly} (to follow inclusive teaching practices that support access) \cite{tac-teaching-inclusively}. Prior resources exist to support computing instructors to teach accessibly \cite{tac-teaching-inclusively, teachaccess, accesscomputing}. Instructors have held Birds of a Feather and workshop sessions aimed at improving and integrating accessibility into computing courses \cite{disability-accessibility-bof-25, disability-accessibility-bof-24, hands-on-workshop}. However, compared to instructors, TAs play a different role and require different skills and training. While computing TAs receive training on skills like teaching techniques \cite{uta-in-cs-lit-review}, we are not aware of any large-scale accessibility trainings specifically designed for TAs in computing education. Others have introduced accessibility training for TAs in non-computing subjects (within a workshop context) with positive results \cite{preparing-tas, accessibility-in-ta-training}.

\section{Auditing and Remediation Systems}
Accessibility audits provide a structured mechanism for identifying access barriers across a course and its digital materials \cite{auditing-electronic-resources}. Remediation refers to the process of addressing accessibility issues in existing content and resources. The first author conducted audits of assignments and materials used in two introductory programming courses, CSE 122\footnote{\url{https://courses.cs.washington.edu/courses/cse122/}} and CSE 123\footnote{\url{https://courses.cs.washington.edu/courses/cse123/}}, including readings and assignment specifications hosted on Ed\footnote{Ed Discussion. \url{https://edstem.org}}, videos, and PowerPoint and Google Slides presentations. We highlight the course structure and auditing for these courses from Spring 2025 in Table~\ref{tab:summary}. The following subsections outline our auditing and remediation strategies across our course sequence.

\begin{table}[ht]
\centering
\small
\caption{Summary of auditing for CSE 122 and CSE 123}
\begin{tabularx}{\linewidth}{l>{\centering\arraybackslash}X>{\centering\arraybackslash}X}
\toprule
& \textbf{CSE 122} & \textbf{CSE 123}\\
\midrule
Lectures/PowerPoints & 21 & 20 \\
Recitation sections & 16 & 16 \\
Assignments & 8 & 8 \\
Unique resources & 83 & 103 \\
Issues identified & 354 & 556 \\
\bottomrule
\end{tabularx}
\label{tab:summary}
\end{table}

\subsection{Classifying Accessibility Issues}
\label{sec:classify}
To better understand the scope and impact of accessibility barriers within our courses, we classified issues across the following dimensions, using templates available in our accompanying website\footnotemark[1].

\begin{itemize}
    \item \textbf{Platform:} The medium in which the issue was found: Ed, section slides, or lecture PowerPoint presentations.
    \item \textbf{Course Element:} The course element (\textit{e.g., }  reading, lesson, assignment) in which the issue appeared.
    \item \textbf{Format:} The format the issue appeared in, such as Video, Text, Image, Animated GIF, Drawing, and Shapes/Objects.
    \item \textbf{Reference:} The specific location of the issue (\textit{e.g.,} ``lecture 7, slide 18'').
    \item \textbf{Context/Purpose:} The content's specific purpose within the course, such as a drawing \textit{illustrating the mechanics of pointer semantics}. 
    \item \textbf{Issue Description:} A detailed description of the issue and obstacles to remediation.
    \item \textbf{Instructional Necessity:} Whether content is required or is not instructionally necessary. 
    \begin{itemize}
        \item \textit{e.g., }  an image showing the contents of a map data structure is required--a student relies on the image to understand its elements and organization; a picture of a corgi in a presentation is likely decorative, and is not instructionally necessary. 
    \end{itemize}
    \item \textbf{Fix suggestion:} A concrete and actionable explanation of how the issue can be remediated.
    \item \textbf{Trivial fix?:} How simple and straightforward the fix is.
    \begin{itemize}
        \item \textit{e.g., } adding alt text to a simple image or replacing a screenshot of code with the code itself would be trivial fixes; re-recording a video or developing a new assignment would be non-trivial fixes.
    \end{itemize} 
\end{itemize}

\subsection{Auditing by Content Type}
We categorized digital course content into four types: text content; images, illustrations, and figures; videos; and presentation materials. We evaluated whether content was Perceivable, Operable, Understandable, and Robust (POUR) \cite{wcag-pour}, mirroring version 2.1 of the Web Content Accessibility Guidelines (WCAG) \cite{wcag}. For example, students who access content across these types nonvisually (\textit{e.g.,} through screen readers or braille) should receive content and contextual information equivalent to that available to sighted peers. Additionally, we recognize the limitations of WCAG 2.1, which does not fully address all disabilities and limits our scope of auditing and remediation.

\subsubsection{Auditing Text Content (\textit{e.g., }  syllabi, readings, webpages)}
\label{sec:audit-text}

Text content can take many formats and modes, introducing a range of accessibility challenges. The most common issues we found were:

\begin{itemize}
    \item Improper usage of headings and other semantic elements, interfering with how AT users navigate documents
    \item Ambiguous or missing hyperlink text (\textit{e.g.,} a raw URL or ``click here''), obscuring context and destination
    \item Text with poor color contrast; evaluated using WCAG 2.1 Level AA Success Criterion 1.4.3 Contrast (Minimum)
    \item Assumptions of visual perception in the language used (\textit{e.g., }  ``as you can see,'' or ``note the highlighted section'')
\end{itemize}

\subsubsection{Auditing Images, Illustrations, and Figures (\textit{e.g., }  diagrams, charts, graphs)}
\label{sec:audit-visual}

Our courses had many essential visual elements that lacked an equivalent nonvisual representation. One common alternative representation is ``alt text'', a concise description used by a screen reader in place of non-text content. Other more complex alternatives also exist, including structured text or data and interactive systems. The most common issues we found were:

\begin{itemize}
    \item Missing or insufficient alt text for images or other non-text content, including hand-drawn diagrams
    \item Use of images to represent text, including code screenshots
    \item Use of text to represent figures, including ASCII diagrams
    \item Visual elements with poor color contrast
\end{itemize}

\subsubsection{Auditing Video (\textit{e.g., }  lecture recordings, explanatory tutorials)}

Videos combine visual and audio elements. In addition to the issues discussed in \ref{sec:audit-text} and \ref{sec:audit-visual}, content should be accessible to Deaf and Hard of Hearing (DHH) students through non-audio representations such as captions or transcripts. The most common issues we found were:

\begin{itemize}
    \item Missing or inaccurate auto-generated captions
    \item Solely relying on drawings or visual demonstrations 
    \item Lack of narration for visual-only cues and actions (\textit{e.g., }  cursor movement or highlighting a portion of the screen)
\end{itemize}

\subsubsection{Auditing Presentation Materials (\textit{e.g., }  slide decks)}

Similarly to videos, presentation materials share issues from \ref{sec:audit-text} and \ref{sec:audit-visual}. However, presentations introduce additional unique challenges due to their fixed visual layouts and implicit sequencing. Our audits focused on presentation source files (further discussed in \ref{presentation-slides}). The most common issues we found were: 

\begin{itemize}
    \item Improper reading order of slide elements, conveying content to screen reader users in the incorrect order
    \item Invisible or off-slide elements which may confuse screen reader users
    \item Grouped items lacking alt text or meaningful structure
\end{itemize}

\subsection{Remediating by Content Type}

We developed content type-specific strategies to remediate issues identified in our audits. At the time of publication, we have started and are continuing to implement these remediation efforts.

\subsubsection{Remediating Text Content}

We rewrote and restructured text to have proper semantics, structure, and use language that is accessible to a wider variety of learners, including those using ATs to interact with content. Our most common remediation strategies were:

\begin{itemize}
    \item Replacing ambiguous link text with descriptive text that communicates the link destination and interaction purpose 
    \item Restyling text to have sufficient color contrast and size
    \item Revising language that assumes a reader's sensory or interaction capabilities (\textit{e.g., }  ``You can see below'')
\end{itemize}

\subsubsection{Remediating Images, Illustrations, and Figures}
We remediated images, illustrations, and figures by focusing on providing contextual information through accessible alternative representations. Our most common remediation strategies included:

\begin{itemize}
    \item Providing text or structured representations (\textit{e.g., }  tables or lists) for hand-drawn diagrams and instructional visuals
    \item Redesigning visual content to have sufficient color contrast
    \item Writing meaningful alt text that is both descriptive and succinct, tailored to the image’s purpose
    \item Using nearby text or captions rather than embedding lengthy alt text for an image, which can disrupt ATs' reading flow 
\end{itemize}

\subsubsection{Remediating Video}

Video remediation was challenging due to its combination of visual, auditory, and temporal elements. Most identified issues were labeled as ``Non-Trivial Fixes'' (as described in Section \ref{sec:classify}) and required substantial editing or a complete re-recording to fix. Our common remediation strategies included: 

\begin{itemize}
    \item Providing equivalent access to content shown visually in the video (\textit{e.g., }  code, slides, diagrams)
    \item Manually writing or correcting auto-generated captions
    \item Verbally narrating on-screen actions/behaviors such as input interactions, window contents, and cursor movement
\end{itemize}

\subsubsection{Remediating Presentation Materials}

We remediated presentation materials at the source level, using tools built into presentation software to ensure that visual and logical structures were aligned and accessible to students using ATs. Our common remediation strategies included: 

\begin{itemize}
    \item Adjusting reading and tab order of elements to follow the desired sequence and expose hidden or invisible elements 
    \item Grouping related shapes and parts together and adding alt text (when relevant) to strengthen context
    \item Providing alt text or captions when content relies on spatial relationships (\textit{e.g., }  highlighted section of a diagram) 
\end{itemize}

\section{Technical Systems}

A purely audit-then-remediate approach is inherently reactive and difficult to scale, as course materials are often updated faster than they can be reviewed and fixed. To address scaling issues, we designed and implemented three different technical approaches for proactive accessibility work. The goal of these systems is \textit{not} to replace manual audits or testing; automated tools lack context and are not comprehensive. Rather, they prevent some accessibility issues and complement existing accessibility workflows.

\subsection{Replacing PDFs with Web-First Materials}

\label{sec:web-first}

PDFs are ubiquitous in delivery of course content. However, the format itself was not designed for accessibility and supports limited accessibility features \cite{scia11y}. Subsequent edits to PDFs generated by software (\textit{e.g.,} a word processor) require re-tagging the entire PDF.

Many PDFs in our courses were primarily structured text, tables, and images --- which could be rewritten as web pages. Web-first materials have major accessibility benefits when compared to PDF and PDF/A \cite{scia11y}. Web accessibility standards are more robust and simpler to teach, web pages better support user customization, and HTML has first-class support for MathML (a language supporting interactive screen reader navigation of math expressions \cite{MathML}).

We tested this approach with reference sheets used in recitation sections by rewriting them in Markdown, compiling them into HTML, and publishing them on our course website. We embedded math using MathJax \cite{MathJax}, a library that converts TeX snippets to MathML. The original PDF could be recovered by using CSS print styles and the ``print to document'' web browser feature.

These updated reference sheets passed the audits we described above. They received significant positive feedback from students and widespread adoption. We expanded this approach to other PDFs in our course, including a course glossary, the final exam reference sheet (also distributed in print), and a seating chart. Designing web-first materials also allowed us to make frequent subsequent edits without needing to re-implement accessible features.

\subsection{Alternate Views for PDF-only Content}

Other content is more challenging to directly redesign as a web page. In these cases, we provided alternate ``views'' into the same content, distributed alongside the PDFs. Importantly, following these steps alone does not make materials accessible; the remediation strategies described in the previous section must also be applied.

\subsubsection{Presentation Slides} \label{presentation-slides} In addition to the previously-mentioned issues, slides exported as PDFs lose animation and object grouping information. To mitigate these issues, we posted source or source-equivalent copies of slides. This also lets users make additional accessibility adjustments (\textit{e.g.,} changing the color or size of a slide element). Both disabled and non-disabled students gave positive feedback, especially for improved search and the ability to edit slides.

\subsubsection{Web Alternatives for TeX-generated PDFs} TeX is often used to author math-related course content. The approaches described in Section \ref{sec:web-first} do not work for complex documents which rely on TeX features that MathJax does not support. Alternate approaches that convert entire TeX source files to HTML fall into two categories:

\begin{enumerate}
    \item tools like pandoc\cite{pandoc} or LaTeXML\cite{LaTeXML} also emulate TeX, but support more TeX macros and packages than MathJax.
    \item tex4ht\cite{TeX4ht} runs a full TeX pass (allowing it to support most macros and packages) and then converts the output to HTML.
\end{enumerate}

Both approaches support MathML. Beyond differences in package support, emulators produce simpler HTML but match original layouts less faithfully. However, tex4ht's complex layouts can be more challenging to navigate with ATs. Further remediation was required after using each tool.

We used pandoc for simple documents without physical layout constraints, such as digital-only worksheets and homework (CSE 331\footnote{\url{https://courses.cs.washington.edu/courses/cse331/25sp/}}). We found tex4ht more effective for complex documents with many macros and layout constraints, such as printed final exams with toggleable solutions.

\subsection{Continuous Automated Testing}

Purely manual accessibility tests scale poorly. Automated accessibility tests catch simple, mechanical issues and let authors focus on and manually test more complex issues. We gradually wrote and adopted two stages of tests: a fast and coarse HTML validator and a more complex automated browser test suite. We ran these locally and in our continuous integration and deployment (CI/CD) pipelines, which provided instant feedback and enforced consistent standards across different iterations of the course staff.

\subsubsection{Lightweight HTML Validation} Valid HTML is a necessary (but not sufficient) condition for accessible web pages. For example, images missing alt text, pages missing a region/landmark, or improper heading structure are invalid HTML and \textit{also} accessibility issues. Invalid HTML is rather common: the WebAIM Million project's 2025 survey of the top million home pages reported that 18.5\% of images lacked alt text \cite{WebAimMillion2025}, while 39\% of pages had improper heading structure \cite{WebAimMillion2025} --- both issues that can be flagged by an HTML validator. We used the Nu HTML Checker \cite{NuHTMLChecker}, resolving all flagged issues, adding it to our CI/CD pipeline, and training staff to use it.

\subsubsection{Automated Browser Testing} Tests that run in-browser catch more issues than validators, such as low-contrast text (found on 79.1\% of WebAIM Million pages\cite{WebAimMillion2025}). Axe-core is an open-source accessibility test framework with specific rules for WCAG 2.1 AA and AAA\cite{Axe}. We wrote automated browser tests with axe-core and used them across several course and research group websites, catching and resolving over 1000 unique issues across over 300 pages (with no false positives). The most common issues that were missed by the validator but flagged by axe-core were low color contrast (especially with syntax highlighting), links without discernible text, issues with page landmarks, and improper use of tables.

Notably, these tests still miss key accessibility issues, including issues that depend on context or human understanding (\textit{e.g.,} if alt text or reading order is \textit{correct}) and specific cases that are challenging to analyze (\textit{e.g.,} text over a background image or gradient). Manual analysis with a human in the loop is highly necessary.

\section{Human Systems}

While the previous systems provide a foundation for improving accessibility in computing education, their impact is limited without broad understanding and adoption. Human systems refer to pedagogical, organizational, and training-based practices that promote accessibility literacy. Next, we describe human systems supporting the adoption of accessibility awareness and practice at scale.

\subsection{Engaging Teaching Assistants}
Our courses rely on both graduate and undergraduate TAs, who often interact more frequently and directly with students than instructors. Thus, it is crucial to engage TAs in accessibility work.

\subsubsection{Activities}

We developed several activities for TAs to reflect on access issues and directly apply accessibility techniques. These activities are also available on our accompanying website\footnotemark[1].

\begin{itemize}
    \item \textbf{``Guess the Assignment''}: TAs listened to the presenter use a screen reader to navigate an unremediated assignment from their course. Despite being familiar with the assignments, TAs struggled to recognize them due to the document being inaccessible: semantic elements were misused and diagrams lacked alt text (without alt text Ed, inserts autogenerated file names such as \texttt{59DLABynUwR0QimwfHHCIc0W}). This activity highlights limitations screen reader users face when content lacks semantic structure and descriptive alternatives. 
    
    \item \textbf{``Spot \& Improve''}: We developed hypothetical vignettes that described real-world accessibility issues in different course settings. We asked participants to read them individually and then discuss in groups  strategies they would use when encountering each situation. Examples included:

    \begin{itemize}
        \item \textit{``During section, while you’re going through review slides on binary trees, a student raises their hand, saying `I can’t really see what you’re pointing at.' You go ahead and increase the size of the slides by zooming in, asking them if it helps. `Kinda, but not really.'''}
        \item \textit{``Office hours are busy—the room is loud, chaotic, and crowded. You start your shift, taking the first student in the queue. They look stressed and overwhelmed. As you sit down, they quietly mention: `Sorry, I have sensory overload, but I really need help with this assignment due tonight.'\text{''}}
    \end{itemize}

    \item \textbf{``Should it alt text?''}:
    We presented TAs with several images across different computing topics and asked if each image is best represented by short, concise alt text or a more complex alternative representation. The core emphasis was that alt text is not always the best solution. For instance, code is likely best represented as text, while complex diagrams such as cache hierarchies or neural networks may be better represented as structured tables or ordered lists.

    \item \textbf{``Fix the Deck''}: TAs were presented an inaccessible slide deck modeled after actual TA-created content: issues included low color contrast, missing or non-descriptive alt text, non-descriptive link text, and improper reading order. TAs were asked to identify at least three distinct accessibility issues and fully remediate at least one issue across all sides.
\end{itemize}

\subsubsection{Trainings} 

We developed and presented trainings using a mix of these activities, in groups as small as 6 and as large as 100. Trainings interspersed explanations of accessibility issues (\textit{e.g.,} images lacking alt text), and relevant laws and standards (\textit{e.g.,} WCAG 2.1 Level AA, Title II of the ADA \cite{ada-fact-sheet}), actionable steps on how to remediate issues, hands-on activities to practice these skills (\textit{e.g.,} ``Fix the Deck''), and group discussions on common patterns and strategies.

Most trainings were in-person and took between 30 minutes and one hour. However, these activities are designed to be flexible: we have adapted them for individual and small group settings, in-person and virtual formats, and for use with TAs and instructors.
 
\subsubsection{Course Staff Accessibility Norms} After these trainings, we set corresponding accessibility expectations for TAs:

\begin{itemize}
    \item \textbf{Avoid images of text.} If the underlying content is text, provide it directly (\textit{e.g.,} code or excerpts of documents).

    \item \textbf{Provide alternative descriptions for necessary images.} Use alt text for images that can be described concisely, and captions or structured text for more complex descriptions.

    \item \textbf{Use structured markup for structured content.} When appropriate, use structured text, headings, tables, and lists.
   
    \item \textbf{Use accessible slide deck practices.} We developed an accessible theme for Google Slides which include colors, typography, and layouts that are WCAG 2.1 AA compliant.
\end{itemize}

\subsection{Engaging Students}

Our courses introduce accessibility to students early and often, through technical and reflective assignments and course accessibility norms. Accessibility is a part of one of our course's broader learning objectives: teaching responsible computing. Teaching students about accessibility has two additional effects: students are active producers of course content (\textit{e.g.,} via discussion board posts), and TAs must take our courses before being hired, and thus join our course staff with a stronger foundation in accessibility.

\subsubsection{Accessibility-Oriented Content} 

Throughout the introductory courses, students engage with assignments and reflections that discuss accessibility in varied contexts. In one project, students debug a program that prints ASCII art, then create additional ASCII designs and write descriptive captions for them. Next, they watch a video featuring a blind software developer sharing how they code and barriers they faced learning to program. Students then reflect on whether their ASCII art and caption would be accessible to a blind user. Other modules explore accessibility in game design or implement a color contrast checker (inspired by \textit{Teaching Accessible Computing} \cite{tac-cs1}). More broadly, students are consistently asked to think critically about how disabled users may use their software.

\subsubsection{Student-Facing Accessibility Norms}

We discouraged the use of code screenshots and explained their accessibility implications. Students who posted questions with a code screenshot were asked to replace it with the text (formatted as a code block) before the course staff answered their question. This led to a reduction code screenshots and prompted positive reflections from students.

\section{Discussion}
In this section, we reflect on the impact and challenges of scaling accessibility efforts within a large introductory programming sequence, and detail limitations to our analysis and future work.

\subsection{Successes}

\subsubsection{More Accessible Materials} We saw anecdotal and quantitative improvements in certain accessibility metrics over time, especially in areas emphasized in training. We noted particular improvements in materials created after our trainings. For example, a brand-new assignment had 22 fewer audited issues than the assignment it replaced \textit{before} any remediation took place, despite having a longer specification and otherwise similar diagrams and math.

\subsubsection{Sustained Engagement from TAs} We received unanimously positive feedback from TAs on the accessibility trainings. After the trainings, some TAs took on additional accessibility tasks within the course, while others engaged in broader campus initiatives, including further trainings and accessibility-related coursework. Several TAs also continued this work in other classes they supported. 

\subsubsection{Sustained Engagement from Students} Students responded positively to assignments and materials that emphasized accessibility. In  reflections and course evaluations, they most often appreciated learning more about ATs and how to use accessibility features when authoring content. For example, students discussed adding alt text to images they uploaded on social media. Students also reflected on how accessibility features supported all users, \textit{e.g.,} by improving legibility, clarity, or search.

\subsubsection{Accommodations Met by Default} When we received disability accommodation requests, many of the requested accommodations related to course materials had already been addressed by our systems. While additional work was still needed, this substantially reduced the scope of necessary remediation.

\subsection{Challenges}

\subsubsection{Gaps in Training} Due to time constraints, trainings emphasized certain issues over others. For example, alt text and color contrast had dedicated activities, while reading and tab order were only briefly discussed. A one-year longitudinal audit saw an \textit{increase} in reading and tab order issues within presentation slides. We attribute this to the lack of hands-on training for reading and tab order and a lack of automated tests for slides.

\subsubsection{Moving Targets} Term-by-term rotation of instructors and staff led to some inconsistencies in accessibility implementation, and resulting changes in material (such as topic order and significant changes to assessments) made drawing direct comparisons of accessibility across terms challenging. 



\subsection{Limitations}

\subsubsection{Non-Comprehensive Audits} Our audits were not intended to demonstrate full compliance with WCAG 2.1 AA, the ADA, or other relevant standards or regulations. We intentionally prioritized issues specific to our materials and remediation strategies that could be executed by TAs with limited prior accessibility knowledge.

\subsubsection{Focus on Course-Authored Materials} Our processes only focused on materials authored by course staff. This excludes textbooks (which our course did not have), external readings and videos, and supporting software (\textit{e.g.,} development environments). Other elements of courses also have major impacts on accessibility, including facilities, course policies, and institutional support.

\subsubsection{Focus on Introductory Computing} This work may not generalize to advanced computing courses, particularly those with unorthodox math or significant physical, graphical, or audio needs.

\subsection{Future Work}

\subsubsection{Broader Accessibility Coverage} These systems address a deliberate subset of access issues--those with frequent accommodation needs, well-documented solutions, and areas where we had specific expertise. We plan to expand our approach to additional accessibility needs, such as audio description, reduced motion, or dyslexia and dyscalculia. Prior work in accessible CS education mainly centers on BLV and DHH students, with less attention to other disabilities and neurodivergent learners. We aim to broaden our strategies to better support these and other underrepresented groups.


\subsubsection{Domain-specific Accessibility Issues} There are many domain-specific accessibility issues in computing, such as how to convey the structure of a complex data structure (\textit{e.g., }  a graph or automata). We hope to scale existing solutions to these challenges, such as work by \citet{GSK} on GSK, an accessible graph sketching tool.

\subsubsection{Improved Longitudinal Auditing}

We intend to continue auditing accessibility of materials in subsequent course offerings. We plan to design improved mechanisms for longitudinal evaluations to account for variability in course offerings and materials.




\subsubsection{Engaging with Broader Communities} We conducted this work mostly within our own institution and discipline. We would like to collaborate with other institutions and in areas beyond computing education, both to share our own materials and learn best practices from others. We are also interested in partnering with organizations that represent disabled students, programmers, and educators.

\section{Conclusion}

We explored how accessibility can be integrated and scaled across an introductory computing sequence through interrelated systems. We discussed how we audited and remediated accessibility issues, then turned to building technical and human systems to support accessibility work at scale. This approach was successful in building accessible materials and engaging with students and staff. However, we faced challenges with gaps in our accessibility trainings, changes to our staff and materials, and limitations in our ability to comprehensively audit content or generalize beyond our initial scope. Our work demonstrates that complex accessibility efforts can be scaffolded to support instructors, TAs, and students in large computing courses. We encourage computing educators to consider building auditing, remediation, technical, and human systems to support sustained, evolving accessibility practices at scale.

\begin{acks}
We thank the UW CSE 12X introductory programming instructors, teaching assistants, and students. We are especially grateful to Jennifer Mankoff, Brianna Wimer, Ather Sharif, Venkatesh Potluri, Kelly Avery Mack, Shuxu Huffman, Ruth Anderson, Gaby de Jongh, Sean Mealin, Justin Hsia, Michael Ball, and Richard E. Ladner, whose input shaped our trainings, tooling, and writing. This work was supported in part by the NSF BPC Alliance AccessComputing (2137312) and the Paul G. Allen School of Computer Science \& Engineering.

\end{acks}

\newpage

\bibliographystyle{ACM-Reference-Format}
\balance
\bibliography{references}

\end{document}